\documentclass[aps,prl,twocolumn,groupedaddress]{revtex4}

\usepackage{graphicx}

\begin{document}

\title{Density-functional studies of tungsten trioxide, tungsten bronzes, and related systems}



\author{B. Ingham$^1$, S.C. Hendy$^{1,2}$, S.V. Chong$^2$ and J.L. Tallon$^{1,2}$}

\address{$^1$Victoria University of Wellington, P.O. Box 600, Wellington, New Zealand}
\address{$^2$Industrial Research Ltd., P.O. Box 31310, Lower Hutt, New Zealand}

\date{\today}

\begin{abstract}

Tungsten trioxide adopts a variety of structures which can be
intercalated with charged species to alter the electronic
properties, thus forming `tungsten bronzes'. Similar optical
effects are observed upon removing oxygen from WO$_3$, although
the electronic properties are slightly different. Here we present
a computational study of cubic and hexagonal alkali bronzes and
examine the effects on cell size and band structure as the size of
the intercalated ion is increased. With the exception of hydrogen
(which is predicted to be unstable as an intercalate), the
behaviour of the bronzes are relatively consistent. NaWO$_3$ is
the most stable of the cubic systems, although in the hexagonal
system the larger ions are more stable. The band structures are
identical, with the intercalated atom donating its single electron
to the tungsten \textsl{5d} valence band. Next, this was extended
to a study of fractional doping in the Na$_x$WO$_3$ system ($0
\leq x \leq 1$). A linear variation in cell parameter, and a
systematic change in the position of the Fermi level up into the
valence band was observed with increasing $x$. In the underdoped
WO$_{3-x}$ system however, the Fermi level undergoes a sudden jump
into the conduction band at around $x = 0.2$. Lastly, three
compounds of a layered WO$_4\cdot\alpha$,$\omega$-diaminoalkane
hybrid series were studied and found to be insulating, with
features in the band structure similar to those of the parent
WO$_3$ compound which relate well to experimental UV-visible
spectroscopy results.

\end{abstract}


\maketitle

\section{Introduction}

Tungsten trioxide has long been studied for its interesting
structural, electronic, and electrochromic properties. Tungsten
trioxide is most stable in a pseudo-cubic, distorted ReO$_3$
structure \cite{90} but can also form metastable hexagonal
\cite{106}, tetragonal \cite{15} and pyrochlore \cite{22, 507}
structures. For all of these structures it is possible to
intercalate mono- or di-valent cations in the vacancies or
channels within the structures to form the so-called `tungsten
bronzes' \cite{15, 17, 151, 163}. The resulting materials exhibit
a continuous colour change \cite{12} and often a large increase in
the electrical conductivity, which becomes metallic in nature
\cite{9}. The most-studied system is that of the cubic alkali
bronzes (M$_x$WO$_3$, M = Group I alkali ion, 0 $\leq$ x $\leq$
1). Of these, the sodium bronze is the only compound that has been
reported with $x=1$ \cite{12}, however in the hexagonal system,
all alkali ions (Li - Cs) have been intercalated to varying
degrees \cite{90}. In the cubic system, even low levels of doping
($x<0.2$) cause a dramatic colour change, which has led to
tungsten trioxide being used as a material in electrochromic
windows \cite{200, 182, 142}. Bulk superconductivity has been
observed at 2-3 K in Na$_{0.2}$WO$_3$ \cite{197}. In addition to
electronic doping, tungsten trioxide can also exhibit
electron-doping via the removal of oxygen, with similar
colouration effects \cite{153}. However colouration and conduction
are not strictly related, as the conductivity in a thin film of
oxygen-deficient WO$_{3-x}$ is much lower than in a thin film of
M$_x$WO$_3$ of the same colour \cite{169}.  This indicates that
the mechanism for mobility is not simply due to delocalised
electrons or holes, but rather changes in the band structure near
the Fermi level, which are different for the two types of material
\cite{169, 365}. Recent work by our group has concentrated on
developing organic-inorganic hybrid materials based on
two-dimensional tungsten oxide sheets (formed by corner-shared
WO$_6$ octahedra) linked with organic diamine molecules \cite{us1,
us2}. The electronic structure of these materials will determine
how they may be used in various electronic applications. This work
presents \textit{ab} \textit{initio} computations on a variety of
tungsten oxide derivatives, including a series of cubic and
hexagonal alkali tungsten bronzes, variable doping Na$_x$WO$_3$
($0 \leq x \leq 1$), oxygen-deficient WO$_{3-x}$, and some simple
layered organic-inorganic tungsten oxide hybrid structures to aid
in understanding the optical and electronic properties of the
latter. During the course of completing this work another paper
has recently appeared which treats some of the earlier materials
\cite{521}.

\section{Computational details}

We have applied density functional theory (DFT) within the
generalized gradient approximation (GGA) \cite{PBE96} using the
VASP package \cite{vasp1,vasp2,vasp3,vasp4} which solves the
DFT-GGA Kohn-Sham equations within the pseudopotential
approximation. Here the valence electrons have been expanded in a
plane wave basis set and the effect of the core on the valence
electrons has been modelled with ultrasoft pseudopotentials. We
used ultrasoft Vanderbilt type pseudopotentials \cite{ultra1} as
supplied by G. Kresse and J. Hafner \cite{ultra2}. The
pseudopotential valence states and cut-off energies for all
elements used are given in Table \ref{table:pp}.

\begin{table*}
\caption{\label{table:pp}Pseudopotential parameters used for all
calculations.}
\begin{ruledtabular}
\begin{tabular}{ccc}
Atom&Valence Electrons&Cut-off energy (eV)\\
\hline
W&$6s^15d^5$&188.192\\
O&$2s^22p^4$&395.994\\
H&$1s^1$&200.000\\
Li&$2s^1$&76.254\\
Na&$3s^1$&48.686\\
K&$4s^1$&70.923\\
Rb&$5s^1$&63.093\\
Cs&$6s^1$&47.697\\
C&$2s^22p^2$&286.744\\
N&$2s^22p^3$&348.097\\
\end{tabular}
\end{ruledtabular}
\end{table*}

To study the effect of fractional doping in the Na$_x$WO$_3$
system, we used a supercell method consisting of up to eight
primitive WO$_3$ cells, with between 1-7 of these cells occupied
by sodium ions in a pseudo-random fashion. Similarly for the
oxygen-deficient WO$_{3-x}$ system, 3-6 cells were used with one
oxygen vacancy in each case. In all cases the basis vectors of the
cell were chosen to avoid the creation of lines or planes of
dopants.

$k$-point meshes between 4$\times$4$\times$4 and
15$\times$15$\times$15 were used to relax the various systems,
corresponding respectively to 24-36 and 120-455 $k$-points in the
irreducible Brillouin zone. As a general rule, the simplest bronze
structures with few (1-4) formula units per cell used finer
$k$-point meshes (more $k$-points) than the larger non-metallic
WO$_{3-x}$ and hybrid structures. In all cases the $k$-point mesh
was varied at the end of the relaxation and the energy was found
to converge to better than 10 meV in all cases. The cell
parameters and atomic positions were allowed to relax in alternate
cycles. The atomic positions were considered relaxed when the
total energy had converged to within 10 meV between ionic steps.
In the calculations of the hybrid structures the relative
positions of the in-plane tungsten and oxygen atoms were fixed at
the origin and half-way along each planar axis, respectively.

\section{Cubic and hexagonal alkali bronzes}

\subsection{Structure}

\begin{figure}
\includegraphics*[width=85mm]{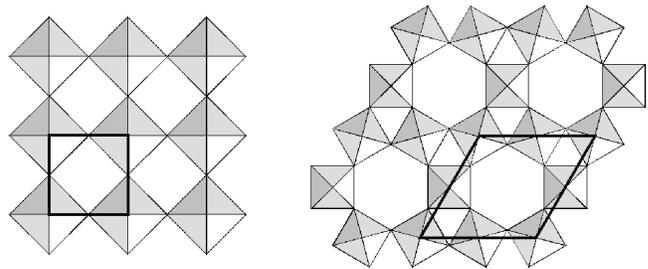}
\caption{\label{fig:hex-cub-struc} The structure of cubic (left)
and hexagonal (right) tungsten oxide and bronzes. The unit cell is
indicated in each case.}
\end{figure}

The alkali elements (H, Li, Na, K, Rb and Cs) were each used as
the intercalated species in cubic MWO$_3$ ($x=1$ bronzes) and
hexagonal M$_{0.33}$WO$_3$ (full intercalation of the hexagonal
tunnels). In each case the intercalated metal atom was placed in
the plane of the apical oxygen atoms, in the centre of the cavity
(Figure \ref{fig:hex-cub-struc}). WO$_3$ does not form a perfectly
cubic cell at room temperature, as the WO$_6$ octahedra are
slightly distorted in terms of W$-$O bond lengths and W$-$O$-$W
bond angles, due to antiferroelectric displacement of the tungsten
atoms and subsequent rotation of the WO$_6$ octahedra
\cite{186,90}. However this is not taken into account in the
calculated system; the cell always relaxes to a cubic structure.
In calculations involving supercells of tungsten trioxide the
distortions are seen \cite{521}. Tungsten bronzes on the other
hand do form the simple cubic structure at high intercalation
levels \cite{10}. The hexagonal structure, which was also studied
for comparison, is the same for both the oxide and bronzes. (The
term `cubic' is used connotatively throughout this article of
those systems that are cubic or close to it; as opposed to the
hexagonal systems also studied.)

\begin{figure}
\includegraphics*[width=85mm]{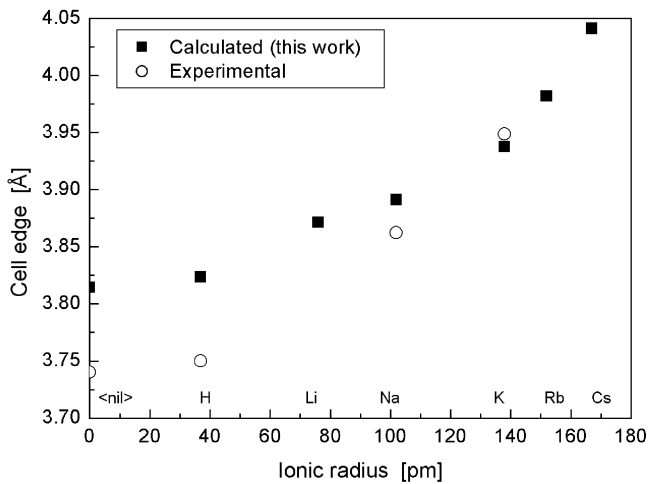}
\caption{\label{fig:cell_cub_MWO3} Calculated and experimental
values of the cell parameter for fully intercalated cubic tungsten
bronzes.}
\end{figure}

Figure \ref{fig:cell_cub_MWO3} shows some experimental results for
the cubic bronze system (obtained from \cite{roth,255,10,chamb}.
In the case of non-cubic WO$_3$ a cubic cell was calculated from
the volume average of the given parameters.) Of all the
intercalated alkali elements which have been attempted
experimentally, only sodium is able to form a stable structure
with $x=1$ at normal temperatures and pressures. In general the
calculated cell parameters for the cubic system are larger than
the experimental. It is also noticeable that as the size of the
intercalated ion increases, the cell size increases
super-linearly. The Goldschmidt tolerance factor for cubic
perovskites can be calculated from the formula $t=\frac{r_{M}
+r_{O}}{\sqrt{2}(r_{W} + R_{O})}$ \cite{gs}, where $r_{j}$ are the
ionic radii. For a perovskite structure to be stable, $t$ must be
less than unity. The tolerance factors for NaWO$_3$ and KWO$_3$
are 0.909 and 1.056 respectively, indicating that the potassium
atom is slightly too large to form a stable structure. Potassium
bronzes have been formed with high $x$ contents, but only under
high-pressure synthesis conditions \cite{chamb}. Rubidium and
caesium cubic bronzes with high $x$ content cannot be formed. The
hexagonal structure, having larger tunnels, is able to accommodate
larger ions than observed in the cubic system. In the experimental
system the stability of hydrogen is notoriously difficult to
maintain, as it is quite mobile due to its small size. Both
hydrogen and lithium are small enough ions to be able to occupy
the small triangular sites between the hexagonal tunnels
\cite{90}. Thus reported experimental results for H$^+$- and
Li$^+$-hexagonal bronzes may show differing behaviour from larger
atoms because the occupied sites may be different in each case. In
the experimental hexagonal system, the $a$ parameter is observed
to increase as ions are intercalated while the $c$ parameter
decreases. The final values are relatively consistent across the
series, as given in Table \ref{table:bronze_cells}. These are
generally lower than the calculated values we have obtained,
although most discrepancies are less than 0.5$\%$.

\begin{table*}
\caption{\label{table:bronze_cells}Experimental cell parameters
for hexagonal tungsten oxide and bronzes (from \cite{90,15}),
compared with our calculated results for hexagonal tungsten
bronzes with hexagonal sites completely occupied. }
\begin{ruledtabular}
\begin{tabular}{ccccc}
Compound&Experimental a ({\AA})&Calculated a ({\AA})&Experimental c ({\AA})&Calculated c ({\AA})\\
\hline
WO$_3$&7.298&7.4103&3.899&3.8144\\
H$_{0.33}$WO$_3$&7.38&7.4173&3.78&3.8111\\
Li$_{0.33}$WO$_3$&7.405&7.4007&3.777&3.8219\\
Na$_{0.33}$WO$_3$&7.38&7.4034&3.775&3.8248\\
K$_{0.33}$WO$_3$&7.37&7.4010&3.77&3.8282\\
Rb$_{0.33}$WO$_3$&7.38&7.4163&3.78&3.8289\\
Cs$_{0.33}$WO$_3$&7.38&7.4507&3.785&3.8342\\
\end{tabular}
\end{ruledtabular}
\end{table*}

For the hexagonal system, the changes in the lattice parameters
are much less pronounced than in the cubic system, as shown in
Table \ref{table:cub-hex-comp}. This is due to the hexagonal
tunnel spaces being much larger than the cavities in the cubic
system and so the interactions between the inserted ion and the
WO$_3$ lattice are smaller.

\begin{table*}
\caption{\label{table:cub-hex-comp}Calculated cell volume per
WO$_3$ unit for cubic and hexagonal tungsten oxide and bronzes. }
\begin{ruledtabular}
\begin{tabular}{ccc}
Compound&Cubic ({\AA}$^3$)&Hexagonal ({\AA}$^3$)\\
\hline
WO$_3$&55.4977&60.4648\\
H$_x$WO$_3$\footnotemark[1]&55.8946&60.5278\\
Li$_x$WO$_3$&58.0127&60.4270\\
Na$_x$WO$_3$&58.9116&60.5158\\
K$_x$WO$_3$&61.0457&60.6785\\
Rb$_x$WO$_3$&63.1294&60.7928\\
Cs$_x$WO$_3$&65.9809&61.4445\\
\end{tabular}
\end{ruledtabular}
\footnotetext[1]{$x=1$ for cubic, $x=0.33$ for hexagonal.}
\end{table*}

\subsection{Charge density}

\begin{figure}
\includegraphics*[width=85mm]{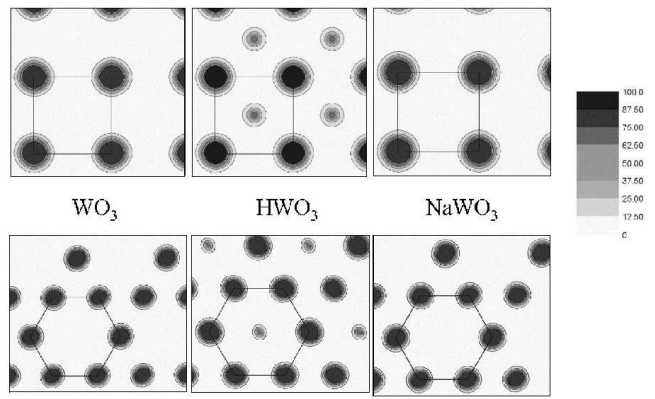}
\caption{\label{fig:cub_hex_chg} Charge density maps of cubic
(top) and hexagonal (bottom) tungsten oxide and bronzes. The
largest tunnels are indicated, which are completely filled in each
case. The oxygen atoms in-plane with the intercalated atom can be
clearly seen, and only in the case of hydrogen is the charge still
associated with the intercalated atom.}
\end{figure}

Charge density plots of the cubic system taken in a plane through
the centre of the cell where the intercalated atom sits, reveal
that hydrogen behaves differently from the other intercalates.
This is shown in Figure \ref{fig:cub_hex_chg}. Bearing in mind
that the charge density plots consider only the valence electrons,
the fact that we observe some charge on the hydrogen atom but not
on any of the others indicates that hydrogen is not ionised. That
larger atoms are completely ionised is consistent with other
reported experimental results \cite{186}. It is interesting that
the same phenomenon occurs in both the cubic and hexagonal cases.
The first ionisation energy of the intercalates are as follows:
hydrogen 1.318 eV, lithium 0.526 eV, sodium 0.502 eV, potassium
0.425 eV, rubidium 0.409 eV, caesium 0.382 eV \cite{SI}. The high
ionisation energy of hydrogen with respect to the other
intercalates may be responsible for the differing behaviour.

\subsection{Energies of formation}

\begin{figure}
\includegraphics*[width=85mm]{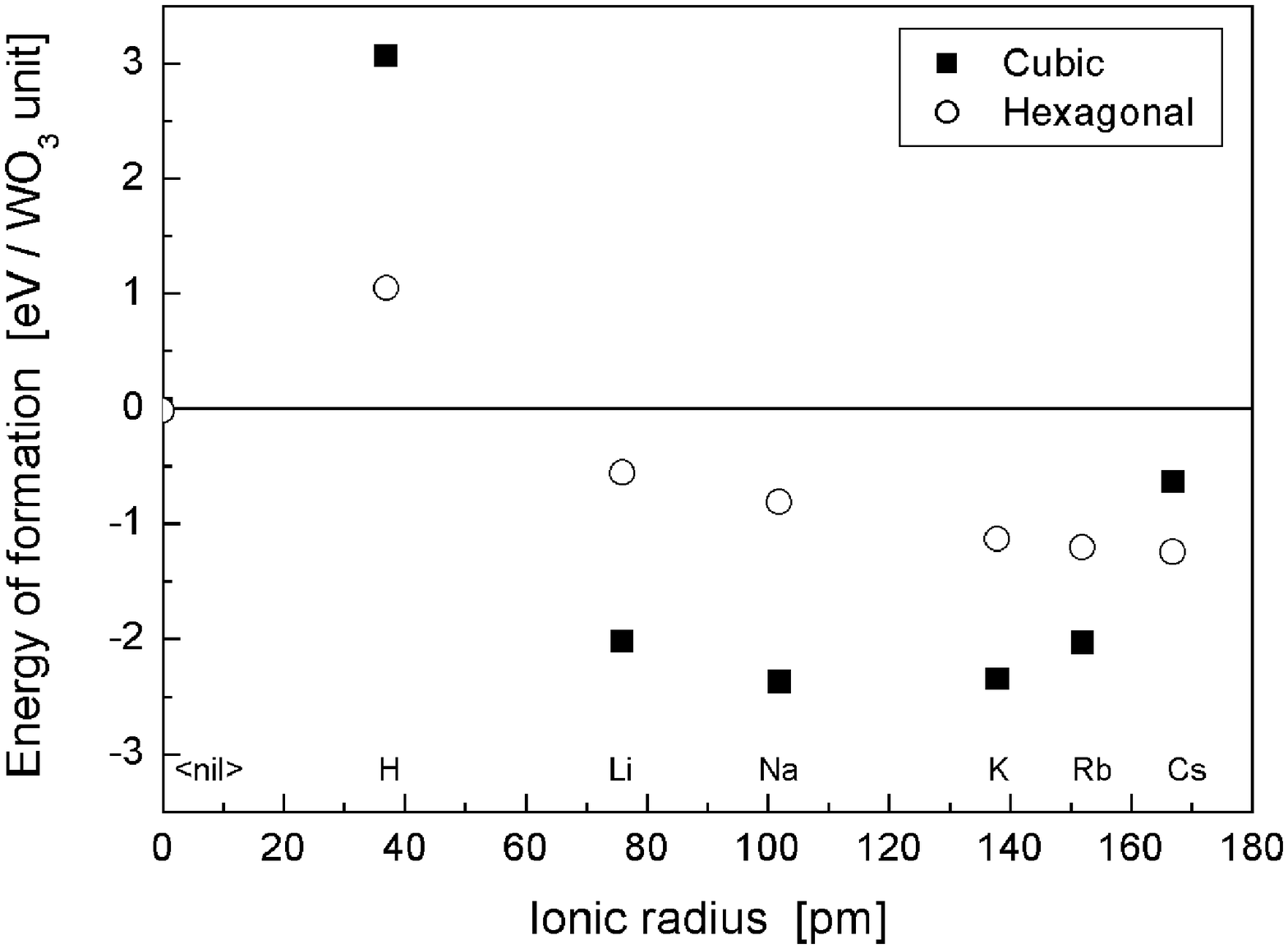}
\caption{\label{fig:bronze_energ} Calculated energies of formation
per WO$_3$ unit for cubic and hexagonal tungsten bronzes, relative
to cubic WO$_3$.}
\end{figure}

The energies of formation of the cubic and hexagonal tungsten
bronzes are shown in Table \ref{table:bronze_energies} and Figure
\ref{fig:bronze_energ}. These are calculated by subtracting the
ground state energies of the components (WO$_3$ plus the metal
cation) from the ground state energy of the final product
(tungsten bronze). A negative energy of formation therefore
indicates that the compound formed is stable.

\begin{table*}
\caption{\label{table:bronze_energies}Calculated energies of
formation per WO$_3$ unit for cubic and hexagonal tungsten oxide
and bronzes, relative to cubic WO$_3$.}
\begin{ruledtabular}
\begin{tabular}{ccc}
Compound&Cubic(eV)&Hexagonal(eV)\\
\hline
WO$_3$&0&-0.016\\
H$_x$WO$_3$\footnotemark[1]&3.059&1.047\\
Li$_x$WO$_3$&-2.023&-0.560\\
Na$_x$WO$_3$&-2.373&-0.813\\
K$_x$WO$_3$&-2.354&-1.132\\
Rb$_x$WO$_3$&-2.039&-1.203\\
Cs$_x$WO$_3$&-0.637&-1.245\\
\end{tabular}
\end{ruledtabular}
\footnotetext[1]{$x=1$ for cubic, $x=0.33$ for hexagonal.}
\end{table*}

Firstly, a comparison of cubic and hexagonal WO$_3$ shows that the
hexagonal phase has a very similar energy to the cubic phase.
Literature results indicate that the cubic phase is preferred,
although the hexagonal phase is stable up to temperatures of
400-500$^o$C, indicating that the phase transition has a high
activation energy \cite{90,106}. In both the cubic and hexagonal
systems, the hydrogen-intercalated bronze energy is positive and
large, indicating that the hydrogen bronzes are not stable. In the
experimental system the hydrogen bronzes are easily oxidised as
the protons are highly mobile \cite{255,hobbs}. This result also
relates to hydrogen being the only intercalate that does not
ionise in the bronze structures, as evidenced by the charge
density plots earlier. For the hexagonal bronzes (apart from
hydrogen) there is a steady downward trend in the energy of
formation as the size of the intercalated alkali metal ion
increases. Therefore the larger ions form more stable bronzes than
the smaller ones, which is also indicated in the literature
\cite{15}. In the cubic system however, the energy drops to a
point and then, beyond Na, increases in the case of the larger
intercalates. This point coincides with the stability predicted by
the Goldschmidt perovskite tolerance factor. For the stable
compounds (WO$_3$, LiWO$_3$, and NaWO$_3$) there is a progressive
decrease in the energy of formation. This supports experimental
evidence that sodium may well be the most stable of the cubic
bronzes, as it is the only one for which a fully intercalated
compound has been reported \cite{hobbs}.

\subsection{Density of states}

\begin{figure}
\includegraphics*[width=85mm]{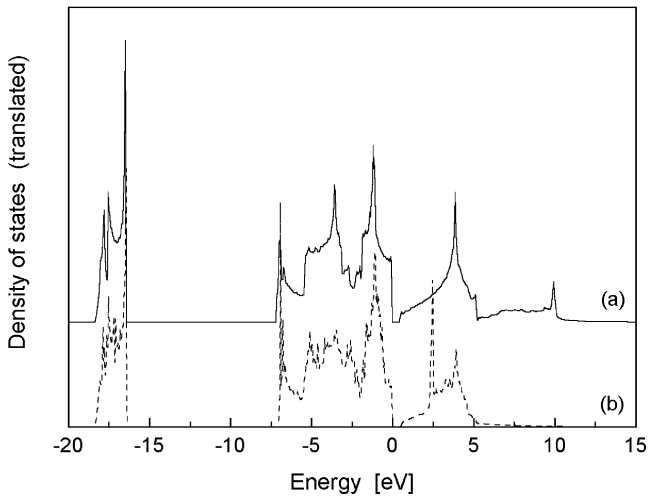}
\caption{\label{fig:WO3_dos_full} The density of states as
calculated for cubic (a) and hexagonal (b) WO$_3$.}
\end{figure}

\begin{figure}
\includegraphics*[width=85mm]{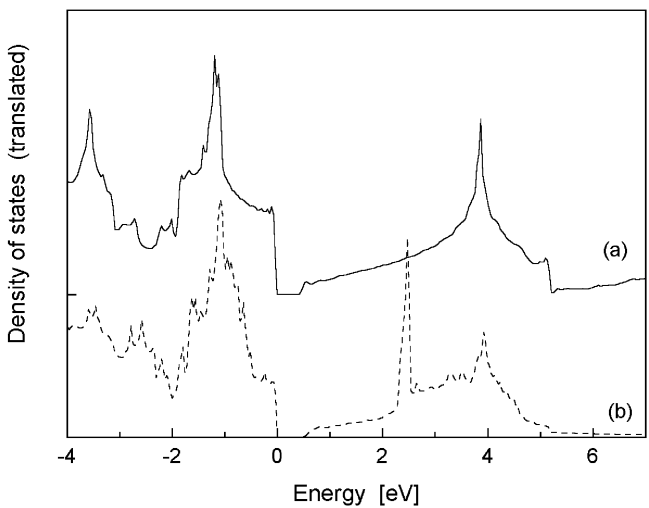}
\caption{\label{fig:WO3_dos_zoom} A detailed view of the
conduction band of cubic (a) and hexagonal (b) WO$_3$, from figure
\ref{fig:WO3_dos_full}.}
\end{figure}

For both the cubic and hexagonal systems other than hydrogen, the
basic band structures of the bronzes are essentially identical to
the parent oxide of the same phase. The only difference amongst
them is the position of the Fermi level relative to the valence
and conduction bands, which will be discussed later. Hence
comparing the band structure of hexagonal and cubic WO$_3$ will
aid a great deal in describing the bronze systems. A comparison of
the density of states for cubic and hexagonal WO$_3$ is shown in
Figures \ref{fig:WO3_dos_full} and \ref{fig:WO3_dos_zoom}. The
lowest band, situated at -18 to -16 eV, corresponds to the oxygen
\textsl{2s} orbitals. This band is present in all of the
tungsten-oxide-based systems studied to date and always occurs at
the same energy regardless of the structure or the presence of
intercalated atoms or molecules.

The broad valence band, from -7 to 0 eV, is comprised mainly of
oxygen \textsl{2p} orbitals. There is a small tungsten \textsl{5d}
component but this is negligible above -2 eV. The valence band
cut-off is sharp at 0 eV and coincides in both the cubic and
hexagonal case with the Fermi level, rendering the oxide materials
semiconducting. The conduction band, detailed in Figure 8, lies
from 0.5 to 5 eV in both the cubic and hexagonal cases. In the
cubic case it is solely comprised of tungsten \textsl{5d}
orbitals; however in the hexagonal case there is also some
additional oxygen \textsl{2p} contribution - particularly to the
strong peak feature observed at 2.5 eV. The band gap (defined as
the difference in energy between the top of the valence band and
the bottom of the conduction band) is 0.4 eV in the cubic system
and 0.5 eV in the hexagonal. This is much less than the observed
band gap, which is typically reported in the range of 2.5 - 3 eV,
and as being indirect \cite{177,182,184}. However this is not too
surprising as DFT generally underestimates band gaps. The presence
of peak features in the density of states can also lead to the
phenomenon where even though the conduction band is being filled,
there is a sudden increase in the population of the conduction
band at these peaks and a sharp transition in the optical spectrum
is observed.

It is worth taking pause here to point out the similarities
between this work and that of experimental results and other
calculations reported on the same structures. X-ray photoelectron
spectroscopy (XPS) reveals that the valence band is comprised of
oxygen \textsl{2p} states only, and the conduction band of
tungsten \textsl{5d} states \cite{182,252}. The oxygen \textsl{2s}
state at $\sim$ -20 eV has also been observed by XPS \cite{180}
although, given the precision of the measurement, this band is
broadened out and appears to extend into the oxygen \textsl{2p}
valence band, the distinction of which is not made by the authors.
Calculations using the local-density approximation and
full-potential linear muffin-tin orbitals (where all electrons are
considered, not just the valence electrons as in the case of the
VASP program) result in density of states spectra which are
virtually identical to those we have obtained \cite{56,111}. This
is the case in both the cubic and hexagonal systems. An older
paper by D.W. Bullet \cite{186}, utilising a non-relativistic
atomic orbital-based method shows a very similar band structure
(oxygen \textsl{2p} as valence band and tungsten \textsl{5d} as
conduction band).

\begin{figure}
\includegraphics*[width=85mm]{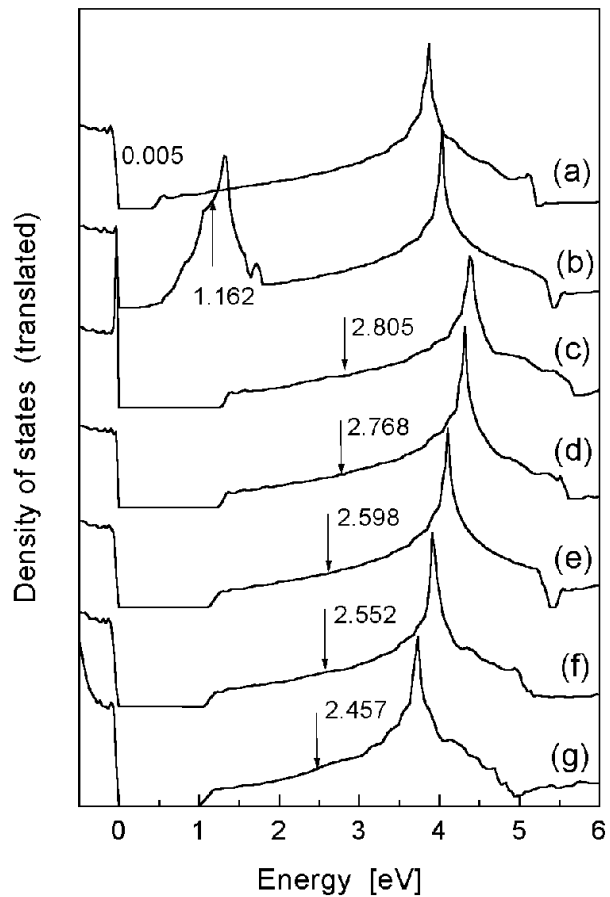}
\caption{\label{fig:dos_cub_MWO3} Calculated density of states for
cubic tungsten bronzes, MWO$_3$, near the Fermi level: (a) WO$_3$,
(b) HWO$_3$, (c) LiWO$_3$, (d) NaWO$_3$, (e) KWO$_3$, (f)
RbWO$_3$, (g) CsWO$_3$. The Fermi level is indicated in each
case.}
\end{figure}

Figure \ref{fig:dos_cub_MWO3} shows a comparison of the density of
states of the cubic bronzes with the parent oxide. The hydrogen
bronze system shows a large peak feature at the bottom of the
conduction band, which is attributed to the non-ionised hydrogen
\textsl{1s} orbital. Because only one electron is contributed from
the hydrogen, the Fermi level lies about in the middle of this
sub-band. For the other bronze systems, analysis of the density of
states contribution from each atom reveals that the intercalated
atom contributes very little if at all to the overall density of
states. This is expected due to its complete ionisation, as the
single valence electron of the intercalated atom is contributed to
the W$-$O framework \cite{186}. All compounds have the Fermi level
located well into the conduction band, rendering them metallic.
The shape of the band structure does not change as atoms are
intercalated. This is also noted in other literature
\cite{56,111}. Also, the magnitude of the band gap and the
position of the Fermi level is relatively constant - even for
those compounds which are known to be unstable (cubic potassium,
rubidium and caesium bronzes).

\begin{figure}
\includegraphics*[width=85mm]{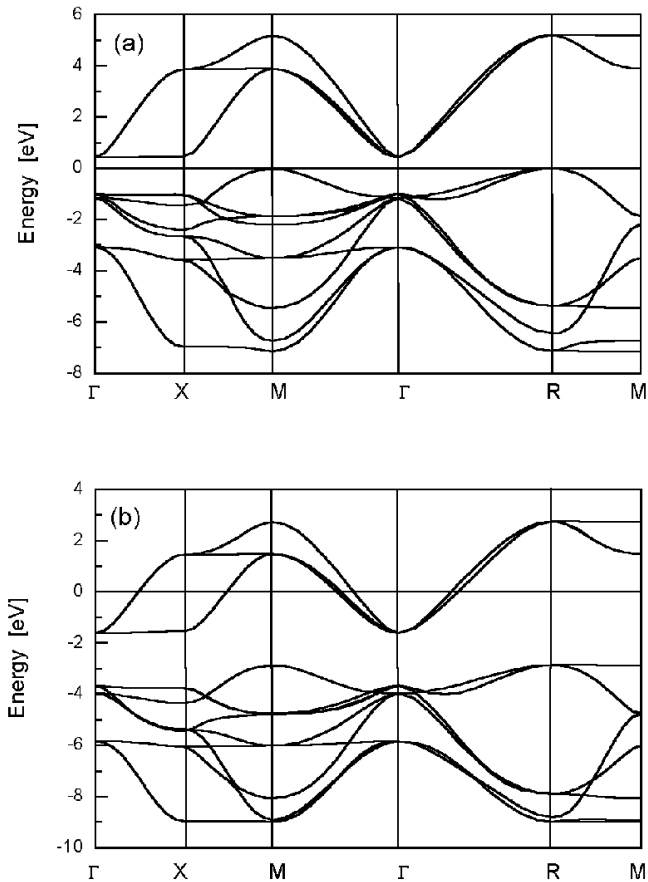}
\caption{\label{fig:Na-WO3-band-comp} Band structure diagrams of
cubic WO$_3$ (a) and NaWO$_3$ (b).}
\end{figure}

We have also derived the band structure curves along lines of
symmetry in cubic WO$_3$ and cubic NaWO$_3$. These are given in
Figure \ref{fig:Na-WO3-band-comp}. These results agree extremely
well with the calculated band structures of both Bullett
\cite{186} and Cora et. al. \cite{111}, considering that different
methods were used for each calculation. We again notice the high
degree of similarity between the parent oxide and the sodium
bronze, with the only major difference being the position of the
Fermi level.

\section{Cubic Sodium Tungsten Bronze Series}

\begin{figure}
\includegraphics*[width=85mm]{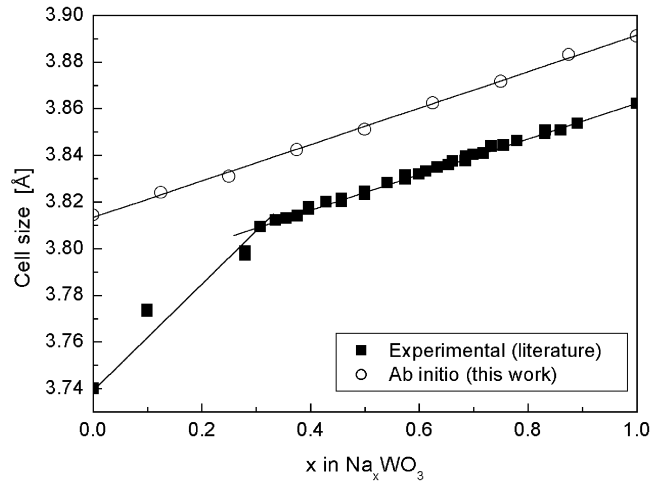}
\caption{\label{fig:NaxWO3_cell} Calculated and experimental
values of cell parameter for cubic Na$_x$WO$_3$ with variable
\textsl{x}.}
\end{figure}

Following the work comparing cubic and hexagonal tungsten bronzes,
we set out to explore the sodium bronze system more thoroughly.
Experimentally it is quite difficult to obtain a completely
saturated sodium tungsten bronze (i.e. $x=1$), and even when an
excess of sodium tungstate is used in the
reaction
$3x$$\cdot$Na$_2$WO$_4$ $+$ $(6 - 4x)$WO$_3$ $+$
$x$$\cdot$W $\rightarrow$ $6$Na$_x$WO$_3$,
 it is not a given that
the resulting bronze will have $x=1$ \cite{10}. There is a raft of
experimental results, however, on sodium bronzes with $x<1$ (refs.
\cite{10,15,8,9,12}, to name but a few), all illustrating that
\textsl{x} is a continuous quantity and not confined to any series
of exactly stoichiometric compounds. In addition to $x=0$ and 1,
which were performed as part of the cubic tungsten oxide and
alkali bronze series, we have calculated the structure and density
of states for $x = \frac{n}{8}; n=1-8$. The average cubic cell
parameter was found to increase linearly with $x$ as shown in
figure \ref{fig:NaxWO3_cell}. Also shown in this figure is a
series of reported experimental results. For higher $x$ contents
the two data sets are very close, with the calculated parameters
being about 0.8$\%$ larger than the experimental; again, a good
result given the approximations made using this method. However
below $x=0.3$, the experimental results differ markedly from the
calculated values, due to a phase change to a tetragonal form at
low $x$ values in the experimental system \cite{11}. The structure
of the tetragonal phase is not related to the cubic phase
\cite{15}. It contains seven WO$_3$ units per cell (28 atoms), and
when one takes the fractional doping into account, the resulting
system quickly becomes much too large to attempt a calculation
with the computational resources available to us. In the work of
Walkingshaw et. al. \cite{521} the volume versus $x$ deviates from
this linear behaviour for $x$ as large as 0.5.

\begin{figure}
\includegraphics*[width=85mm]{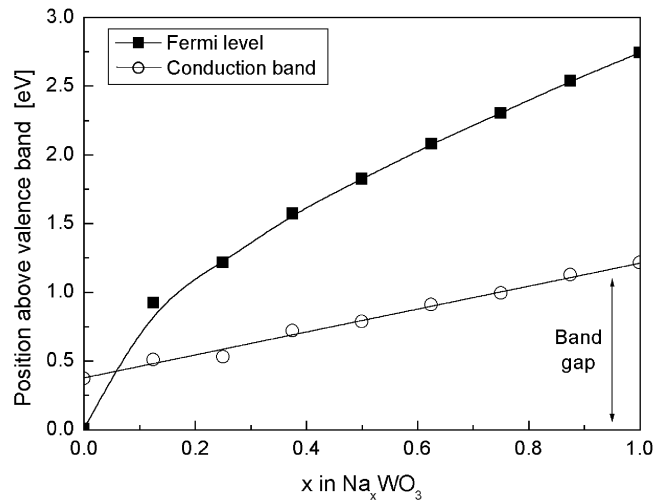}
\caption{\label{fig:NaxWO3_fermi} Band structure of Na$_x$WO$_3$
near the Fermi level showing the movement of the Fermi level into
the conduction band at low $x$ values. The lines are given as a
guide.}
\end{figure}

There are no obvious changes in the appearance of the density of
states as $x$ increases from zero to one. The nature of the band
structure near the Fermi level is shown in figure
\ref{fig:NaxWO3_fermi}. The band gap increases linearly with
\textsl{x}, while the Fermi level quickly moves into the
conduction band. According to this plot we would expect to see a
semiconductor-metal transition at about $x=0.06$, where the Fermi
level moves into the conduction band. However, in the experimental
system this transition is observed at $x=0.3$, corresponding to
(and perhaps influenced by) the structural transition (\cite{15},
and references therein).

Once again, analysis of the individual atomic contributions to the
density of states indicates that each sodium atom is fully
ionised, and the electron donated to the tungsten \textsl{5d}
conduction band \cite{186}.

\section{Sub-stoichiometric WO$_{3-x}$ Series}

As a complement to the sodium bronze series, sub-stoichiometric
WO$_{3-x}$ was examined for compounds with $x$ ranging from zero
to 0.33. In the experimental system, the maximum oxygen loss that
can be achieved without a drastic phase change is $\sim$ 0.35
(\cite{18}, and references therein). There are a number of
different stoichiometric formulae for compounds in the range
WO$_{2.65}$ - WO$_3$, and for some of these the crystal structure
has been solved. They are given the names $\alpha$-, $\beta$- and
$\gamma$-phase, as shown in Table \ref{table:WO3-x_phases} with
their respective formulae. The crystal structures of the $\beta$-
and $\gamma$-phases were solved by Magneli \cite{263,264}. Booth
et al. generalise the $\beta$-phase even further by describing the
existence of crystallographic shear planes \cite{265}. This can
account for the broad range of compositional formulae. While we
are unable to calculate the properties of these phases as
described in literature due to the restriction on the number of
atoms in the system, we are able to observe the effect that
removal of oxygen has on the simple cubic WO$_3$ phase.

\begin{table*}
\caption{\label{table:WO3-x_phases}Compositional ranges for
sub-stoichiometric tungsten oxide species. From \cite{18}.}
\begin{ruledtabular}
\begin{tabular}{cccc}
Phase&Formula&Range&Average\\
\hline
$\alpha$&WO$_3$&WO$_{2.95}$ $-$ WO$_3$&WO$_3$\\
$\beta$&W$_{20}$O$_{58}$&WO$_{2.88}$ $-$ WO$_{2.94}$&WO$_{2.90}$\\
$\gamma$&W$_{18}$O$_{49}$&WO$_{2.65}$ $-$ WO$_{2.76}$&WO$_{2.72}$\\
$\delta$&WO$_2$&WO$_{1.99}$ $-$ WO$_{2.02}$&WO$_2$\\
\end{tabular}
\end{ruledtabular}
\end{table*}

\begin{figure}
\includegraphics*[width=85mm]{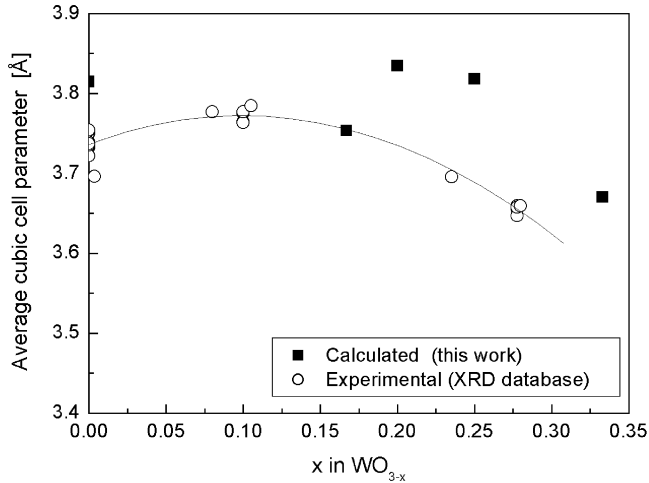}
\caption{\label{fig:WO3-x_cell} The volume-averaged cubic cell
parameter for calculated and experimental sub-stoichiometric
`cubic' WO$_{3-x}$ systems. The curve is given as a guide.}
\end{figure}

As one might expect, removing oxygen from a site causes a local
distortion of atoms around the vacancy, and the cell ceases to be
simple orthorhombic. The cell volume changes; initially there is a
slight increase at low deficiencies, followed by a decrease. These
results are shown in Figure \ref{fig:WO3-x_cell} and are in good
agreement with experimental values, despite the absence of the
phase change in the calculated system.

The energies of formation for the different species studied are
given in Table \ref{table:enform_WO3-x}. We note that a slight
deficit of oxygen ($x=\frac{1}{6}$) is more favourable
energetically than stoichiometric WO$_3$. This is observed
experimentally, as commercial WO$_3$ powder exhibits a loss of
oxygen over 1-2 days in atmospheric conditions. Further loss of
oxygen renders WO$_{3-x}$ less energetically favourable than its
parent oxide, and once again, as was the case with sodium tungsten
bronzes, the presence of a phase change in the experimental system
may explain any discrepancies seen.

\begin{table*}
\caption{\label{table:enform_WO3-x}Energies of formation of the
calculated WO$_{3-x}$ species calculated by the formula
$E_F=E_{TOTAL}-\sum E_{PARTS}=E(WO_{3-x})- (E(WO_3)-\frac{x}{2}
E(O_2)$).}
\begin{ruledtabular}
\begin{tabular}{cc}
Formula&Energy of formation (eV/unit formula)\\
\hline
WO$_3$ ($x=0$)&0\\
WO$_{2.833}$ ($x=\frac{1}{6}$) &-0.103\\
WO$_{2.8}$ ($x=\frac{1}{5}$) &0.447\\
WO$_{2.75}$ ($x=\frac{1}{4}$) &1.007\\
WO$_{2.667}$ ($x=\frac{1}{3}$) &1.661\\
\end{tabular}
\end{ruledtabular}
\end{table*}

\begin{figure}
\includegraphics*[width=85mm]{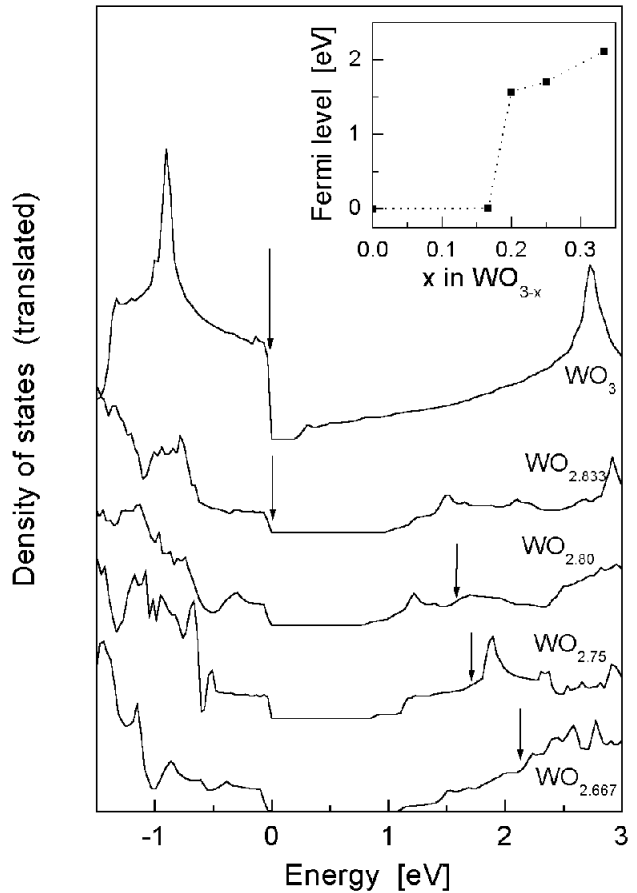}
\caption{\label{fig:WO3-x_dos} Density of states for the
WO$_{3-x}$ system, all with the valence band set at zero. Arrows
show the position of the Fermi level. Inset: Position of the Fermi
level relative to the top of the valence band.}
\end{figure}

It is also of interest to look at the changes in the density of
states as oxygen is removed from WO$_3$. As mentioned in the
introduction, oxygen-deficient WO$_3$ exhibits an increased
conductivity, but not as great as that due solely to the presence
of doped electrons. Figure \ref{fig:WO3-x_dos} shows the density
of states for the WO$_{3-x}$ system in the region near the Fermi
level. The overall spectra share the same features previously
detailed for WO$_3$: the oxygen \textsl{2s} band near -18 eV; the
broad valence band, comprised mainly of oxygen \textsl{2p}
orbitals, from -7 to 0 eV; and the conduction band, consisting
solely of tungsten \textsl{5d} orbitals, lying from roughly 0.5 to
5 eV. Naturally the sub-stoichiometric systems appear more
`jagged' than the parent WO$_3$ compound, due in part to the
breaking of symmetry, rendering each atom non-equivalent to others
within the cell, and causing its contribution to be slightly
different.

The inset of Figure \ref{fig:WO3-x_dos} shows the progression of
the Fermi level into the conduction band. There is a sharp jump
between $x=0.167$ and 0.2 as the Fermi level moves up into the
conduction band - not a gradual transition as in the case of the
sodium bronzes Na$_x$WO$_3$. While the $x=0.167$ compound has the
Fermi level at zero (and therefore, still non-conducting), the
band structure of the valence band is similar to that of the
conducting species. It appears that there is a decrease in the
density of states in the valence band, followed by the Fermi level
being pushed up into the conduction band. The stoichiometry at
which this insulating-conducting transition occurs is in good
agreement with the literature value of WO$_{2.76}$, which
coincides with the $\beta$-$\gamma$ structural phase transition
\cite{365}.

\section{Organic-Inorganic Layered Hybrids}

Following on from the background studies of tungsten bronzes and
the oxygen-deficient tungsten oxides, layered organic-inorganic
hybrid compounds were studied. These structures consist of WO$_4$
layers (as in H$_2$WO$_4$ \cite{512}), connected via aliphatic
(linear) alkyl diamines. These have been investigated
experimentally by us elsewhere \cite{us1}. Three different length
alkyl amines were used in the calculations, with two, four and six
carbons. The input structure of the hybrid systems is the most
conceptually simple: a single unit formula,
WO$_4\cdot$H$_3$N(CH$_2$)$_n$NH$_3$ ($n=2,4,6$ - hereby
abbreviated to W-DAn). It is highly conceivable that the
calculated structure of the hybrid compounds is in a slightly
higher energy state than that of the actual structure, which may
be a supercell of the simple input case, with possible tilts and
rotations of the octahedra and organic molecules. As mentioned,
WO$_3$ does not form a simple cubic structure, but exhibits small
distortions of bond lengths and angles which render it very
slightly off-cubic, with eight formula units per cell. In
extending the computations to the organic-inorganic systems then,
several constraints were necessary. Firstly only one cell was
used, which may affect the outcome not only due to the removal of
distortion in the inorganic layer (which apparently lowers the
energy in the oxide and hydrate compounds) but also because this
does not allow for the differing orientation of the organic
molecules in neighbouring cells. Secondly, constraints were placed
on the inorganic atoms in order to maintain the position of the
layer. This entailed fixing the position of the tungsten and
planar oxygen atoms at the corner and edges of the cell
respectively. This was sufficient to relax the atoms to a sensible
structure.

\begin{figure}
\includegraphics*[width=85mm]{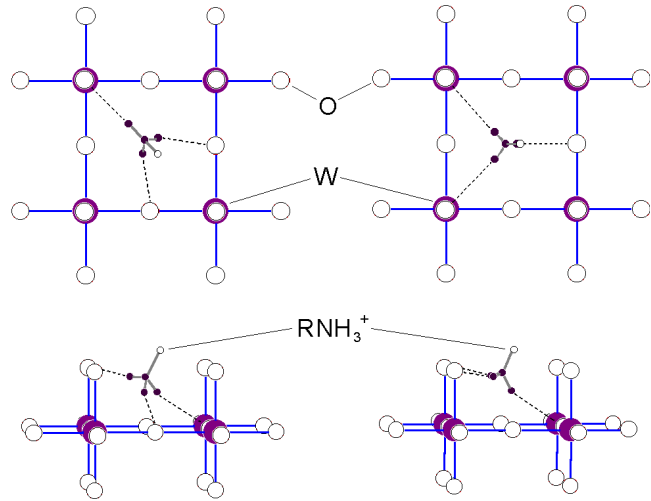}
\caption{\label{fig:term_bridg} Schematic diagrams illustrating
the two bonding configurations, `bridging' (left) and `apical'
(right), in organic-inorganic hybrid materials. Only the first
carbon atom of the organic molecule is shown for simplicity.}
\end{figure}

These compounds are isomorphic with diaminoalkane metal halides
which have been investigated by Mitzi \cite{4}. There, two schemes
were identified for the bonding of the organic ammonium group to
the inorganic layer, designated `bridging' and `terminal' (which
we shall call 'apical'). In the `bridging' case the organic
ammonium forms hydrogen bonds to two bridging and one apical atom
(in the work of Mitzi, this is a halogen atom; in our work,
oxygen) while in the `apical' case hydrogen bonds are formed to
two apical and one bridging atom (Figure \ref{fig:term_bridg}).
This causes the alkyl chain to lie diagonally within the cell when
the ammonium group is in a bridging configuration, but parallel
for the apical. When the alkyl chain is longer than one carbon
(methylamine), the second carbon in the chain would be too close
to the opposing apical (oxygen) atom if a bridging conformation
were adopted. Thus in general, apical bonding is observed for
organic-inorganic systems with organic chain lengths of two or
more carbon atoms.

\begin{figure}
\includegraphics*[width=85mm]{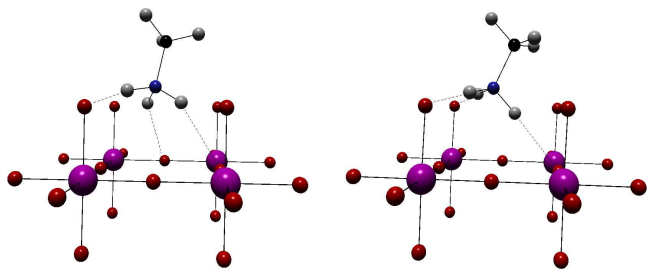}
\caption{\label{fig:W-MA1} Calculated structure of methylamine in
bridging (left) and apical (right) configurations.}
\end{figure}

Given this, calculations were initially performed on the
methylamine system in both the bridging and apical configurations,
to compare the energies of each and as a starting point to
construct the initial cells for the computations of the longer
chain diamine hybrids. Figure \ref{fig:W-MA1} shows the calculated
relaxed structures of one methylammonium ion with respect to the
tungsten oxide layer. In both cases the calculated structure
closely resembles that expected from the diagram in Figure
\ref{fig:term_bridg}. In the apical case the shortest hydrogen
bonds between the ammonium hydrogen atoms and the oxygen atoms of
the inorganic layer are as expected: two short hydrogen bonds to
two apical oxygens and one to the opposing bridging oxygen.
However, for the bridging case the two hydrogens that were
expected to interact with adjacent bridging oxygens are actually
closer to apical oxygens. In this latter case the tilting of the
organic molecule is much less than for the apical case, and
overall it appears that there is a more delocalised attraction
between the hydrogen and oxygen atoms.

\begin{figure}
\includegraphics*[width=85mm]{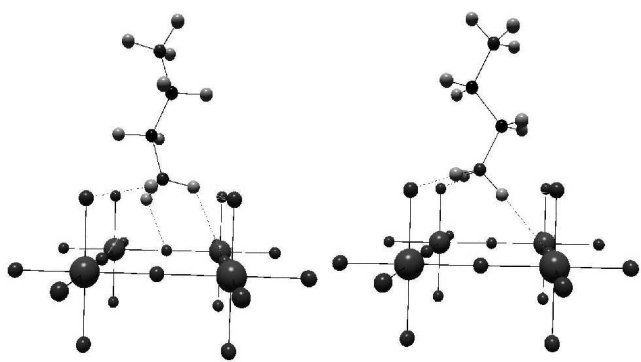}
\caption{\label{fig:W-DA2} Calculated structures of diaminoethane
(DA2) in bridging (left) and apical (right) configurations.}
\end{figure}

The energies of the two systems were calculated to be -123.31 eV
for the apical case and -123.08 eV for the bridging. Thus the
apical configuration appears to be more stable, both from its
lower energy and examination of the structure with more localised
forces. Each of the two methylamine structures were used as a
basis for the initial positions of the diaminoethane (DA2)
compound. The relaxed cell parameters are given in Table
\ref{table:W-DA2}. In the apical case the cell volume is slightly
less than in the bridging case. Despite the greater tilting
expected for the terminal structure, the planar axes ($a$ and $c$)
are shorter and the interlayer spacing $b$ is longer than for the
bridging. The two structures are shown in Figure \ref{fig:W-DA2}.
As for the methylammonium structures, in the bridging conformation
there are several longer bond distances from each hydrogen to the
oxygen atoms, whereas in the apical conformation for each hydrogen
there is a single bond that is distinctly shorter to one oxygen
than the others.

\begin{table*}
\caption{\label{table:W-DA2}Calculated cell parameters of W-DA2
with the organic molecule in the bridging and apical
conformations.}
\begin{ruledtabular}
\begin{tabular}{lcc}
&W-DA2 bridging&W-DA2 apical\\
\hline
a ({\AA})&3.9443&3.9108\\
b ({\AA})&8.7345&8.7992\\
c ({\AA})&3.9443&3.9245\\
$\alpha$&89.98&89.99\\
$\beta$&90.02&90.01\\
$\gamma$&90.02&90.01\\
Volume ({\AA}$^3$)&135.8266&135.0498\\
\end{tabular}
\end{ruledtabular}
\end{table*}

The energies of these two structures are calculated to be
$-114.709$ eV for the bridging conformation and $-115.750$ eV for
the apical. The apical conformation is therefore more stable, and
the difference between the two is greater than for the
methylammonium case. Two other systems were extended from the
apical W-DA2: W-DA4 (4-carbon chain) and W-DA6 (6-carbon chain).
Again the planar tungsten and oxygen atom positions were fixed.
The energies of formation are given in Table \ref{table:W-DAn}. As
can be seen from the negative values, all three compounds are
stable, with W-DA4 being the least stable of the three. This seems
to be confirmed experimentally as W-DA4 is harder to form than
both W-DA2 and W-DA6.

\begin{table*}
\caption{\label{table:W-DAn}Energies of formation of the
calculated W-DAn compounds, calculated by the formula
$E_{F}=E_{TOTAL}-\sum E_{PARTS} = E(WO_{4}\cdot
H_{3}N(CH_{2})_{n}NH_{3})-(E(H_{2}WO_{4})+E(H_{2}N(CH_{2})_{n}NH_{2})$).}
\begin{ruledtabular}
\begin{tabular}{cccc}
Compound&Ground state energy (eV)&Ground state energy of organic molecule&Energy of formation\\
\hline
W-DA2&-115.750&-64.172&-0.939\\
W-DA4&-148.189&-97.506&-0.044\\
W-DA6&-181.887&-130.714&-0.534\\
\end{tabular}
\end{ruledtabular}
\end{table*}

\begin{figure}
\includegraphics*[width=85mm]{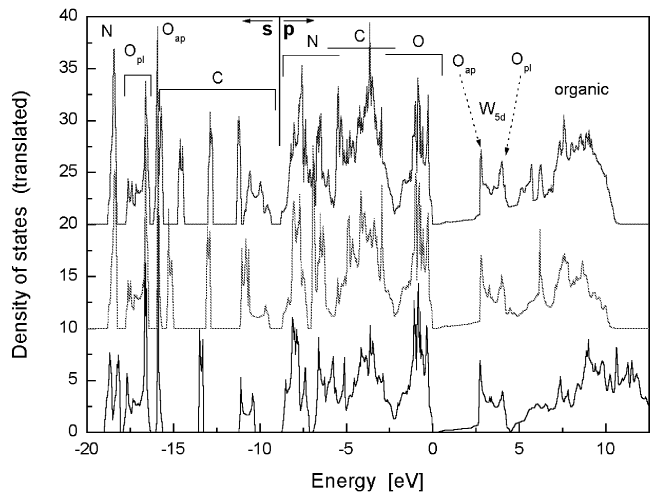}
\caption{\label{fig:W-DAn_DOS} Calculated density of states for
W-DA2 (bottom), W-DA4 (middle) and W-DA6 (top). The Fermi level is
located at E=0.}
\end{figure}

The density of states of the three compounds are all very similar.
The results are shown in Figure \ref{fig:W-DAn_DOS}. The main
features are as follows: As in the tungsten oxide and tungsten
bronze systems, the oxygen \textsl{2s} band is located between -18
and -16 eV. There is a splitting between the planar and apical
oxygen contributions, with the planar oxygen bands being broader
and lying at slightly lower energies. The nitrogen \textsl{2s}
bands lie at about -18.5 eV and the carbon \textsl{2s} bands lie
between -16 and -9 eV. The appearance of multiple carbon s bands
in the longer chain systems is due to the different environments
in which the carbon atoms are located along the length of the
chain. Between -9 and 0 eV lie the \textsl{2p} bands of N
(lowest), C (middle) and O (highest). The hydrogen atoms
associated with the carbon and nitrogen atoms contribute their
single \textsl{1s} electrons to the bands of their respective
atoms.

The oxygen band, from -2 to 0 eV, closely resembles that of the
tungsten oxides and bronzes. There is relatively little organic or
tungsten component to this band, so once again the valence band is
comprised of oxygen \textsl{2p} orbitals. The conduction band
begins at around 0.4 eV but the density of states is very low up
to a peak feature at about 2.7 eV. There is also a second peak
feature at 4.0 eV. This band is comprised mostly of tungsten
\textsl{5d} orbitals, as in the tungsten oxides and bronzes, but
in these hybrid compounds there is an additional oxygen
\textit{2p} component to this band. The apical oxygen atom
contributes to the first peak feature and the planar oxygen atoms
contribute to the second. Above 4.5 eV lie the organic
anti-bonding orbitals. The band structure in the vicinity of the
Fermi level (i.e. valence and conduction bands) is virtually
identical for the three different organic intercalates. As
expected the organic molecule does not participate in electronic
conduction and the compound is an insulator. Features in the
calculated band structure relate very closely to the band gap as
determined by UV-visible spectroscopy \cite{us3}. Powder spectra
of the hybrids indicate a shift in the absorption edge from
2.6-2.8 eV for WO$_3$ to 4.1 eV for the hybrids \cite{us1}. There
was no systematic variation amongst the positions of the
absorption edges for the hybrid materials. It appears likely that
the features in the UV-visible powder spectra correspond to
indirect band gaps, which are displayed in the calculated density
of states spectra as peak features in the valence band, where the
occupancy of the band suddenly increases.

\section{Summary}

A number of tungsten oxide-based systems have been studied using
\textit{ab initio} computation techniques. Hexagonal and cubic
alkali tungsten bronzes exhibit trends in cell sizes which agree
well with experimental data. The band structure and charge density
plots of these show that the intercalated alkali atom (with the
exception of hydrogen) donates its electron to the conduction
band. An in-depth study of the partially doped cubic sodium bronze
system showed the progressive movement of the Fermi level into the
conduction band. However it is suspected that in the experimental
system the onset of metallic conductivity is associated with or
induced by a phase change not accounted for in these calculations.
A transition to a conducting state is also observed experimentally
in oxygen-deficient tungsten trioxide. This is modelled well by
the calculations, as are the changing cell dimensions. The free
energy indicates that a slight deficit of oxygen renders tungsten
trioxide more stable than the exactly stoichiometric form, which
is also observed experimentally. Lastly three
tungsten-oxide/organic hybrids with simple
$\alpha$,$\omega$-diaminoalkane molecules were studied. They are
energetically stable and exhibit many similarities in the band
structure to that of the parent cubic tungsten trioxide. The
amines are protonated to form ammonium groups and the undoped
diammoniumalkane hybrids are calculated to be electrically
insulating.

\section{Acknowledgments}

The authors would like to acknowledge the financial assistance
from the New Zealand Foundation of Research Science and Technology
(Contract number: IRLX0201), The Royal Society of New Zealand
Marsden Fund, and the MacDiarmid Institute for Advanced Materials
and Nanotechnology (Victoria University, New Zealand).

\bibliography{vasp}
\end{document}